\documentstyle[12pt,epsfig]{article}
\newcommand{\be}{\begin{equation}}
\newcommand{\ee}{\end{equation}}
\newcommand{\bea}{\begin{eqnarray}}
\newcommand{\eea}{\end{eqnarray}}
\newcommand{\inn}{\!\cdot\!}
\newcommand{\norm}[1]{\raise.3ex\hbox{:}#1\raise.3ex\hbox{:}}

\begin{document}
\pagestyle{plain}
\setcounter{page}{1}
\newcounter{bean}
\baselineskip16pt


\begin{titlepage}
\begin{flushright}
PUPT-1938\\
HUTP-00/A023\\
hep-th/0006085
\end{flushright}

\vspace{7 mm}

\begin{center}
{\huge 
1+1 Dimensional NCOS\\ and its $U(N)$ Gauge Theory Dual
}

\end{center}
\vspace{10 mm}
\begin{center}
{\large  
Igor R.~Klebanov\\
}
\vspace{3mm}
Joseph Henry Laboratories\\
Princeton University\\
Princeton, New Jersey 08544\\
\vspace{3mm}
{\large  
Juan Maldacena\\
}
\vspace{3mm}
Lyman Laboratories\\
Harvard University\\
Cambridge, MA 02138
\end{center}
\vspace{7mm}
\begin{center}
{\large Abstract}
\end{center}
\noindent

We study some aspects of open string theories on D-branes with critical 
electric fields. 
We show that the massless open string modes that move in the
direction of the electric field decouple. In the 1+1 dimensional case the dual
theory is $U(N)$ SYM with electric flux, and the decoupling of
massless open strings is dual to the decoupling of the $U(1)$ degrees
of freedom. We also show that, if the direction
along the electric field is compact, then
there are finite energy winding closed string modes.
They are dual to Higgs branch excitations of the
SYM theory, and their energetics works accordingly.
These properties provide new non-trivial evidence for the duality.

\vspace{7mm}
\begin{flushleft}
June 2000

\end{flushleft}
\end{titlepage}


\renewcommand{\epsilon}{\varepsilon}
\def\fixit#1{}
\def\comment#1{}
\def\equno#1{(\ref{#1})}
\def\equnos#1{(#1)}
\def\sectno#1{section~\ref{#1}}
\def\figno#1{Fig.~(\ref{#1})}
\def\D#1#2{{\partial #1 \over \partial #2}}
\def\df#1#2{{\displaystyle{#1 \over #2}}}
\def\tf#1#2{{\textstyle{#1 \over #2}}}
\def\Leff{L_{\rm eff}}
\def\at{{\tilde \alpha}}
\def\a{{\alpha}}
\def\pt{{\tilde \psi}}
\def\p{{\psi}}
\def\pa{\partial}
\def\lm{\lambda}
\def\cd{\cal D}
\def\ce{\cal E}
\def\ep{\varepsilon}
\def\cf{{\cal F}}
\def\nn{\nonumber}
\def\cm{{\cal M}}


\section{Introduction}

We study some aspects of non-commutative open string theories. 
These are open string theories on branes in the presence of 
a critical electric field. More precisely, they are defined through
a scaling limit where the electric field is taken to its critical 
value \cite{SST,GMMS} (for earlier work, see \cite{GKP},
and for other recent work \cite{Lenny,Ganor,Barbon,gomis,german}). 
The resulting open string theory is the same as the usual open
string theory but with a new product which includes Moyal phases
in the directions of the electric field, which we take to be
the $01$ directions. The Moyal phase is determined in terms
of the string scale $\alpha_e'$. 

This open string field theory can be defined in various dimensionalities.
We concentrate our discussion on the 1+1 dimensional case. 
In this case the theory on a D1 brane with $N$ units of
electric flux is related to a 
$U(N)$ SYM theory in a sector with one unit of electric 
flux \cite{V,GKP,GMSS}. 
$N$ is related to the open string coupling through the formula 
$ G_0^2 = { 1 \over N}$. Thus, $G_0^2$ cannot exceed $1$ and at this 
maximum value the NCOS actually becomes free \cite{IKtalk}.
To achieve weak string coupling, we need to take $N$ to be large.
In this way, the $1+1$ dimensional NCOS provides a new connection between
large $N$ gauge theory and weakly coupled strings. 
This duality has very interesting 
predictions for the open string field theory. 

\begin{itemize}

\item

The gauge theory dual to the NCOS has gauge group 
$U(N)= U(1) \times SU(N)/Z_N$ theory. This
implies that we have a {\it free} $U(1)$ theory. We stress
that it is free at all energies, not just low energies. 
The $SU(N)/Z_N$ theory with one unit of $Z_N$ flux 
has a mass gap \cite{W}. We show in this paper that 
the massless open strings in 
1+1 dimensional NCOS decouple and give the free $U(1)$ predicted 
by the duality. All other open string modes are massive: they
describe excitations in the $SU(N)/Z_N$ theory.

\item

The field theory has a Higgs branch which has a positive energy 
density. This implies that if we compactify the field theory 
on a circle then we should observe  
a continuum above a certain calculable threshold corresponding 
to breaking $ U(N) \to U(N-1)\times U(1)$ by a Higgs vev. 
The low energy excitations should now reflect the
presence of massless $U(1)$ modes.
We show that the NCOS theory
couples to closed strings that wind around the circle with only
one orientation.
 The spectrum of these closed strings is {\it precisely}
as expected from the duality. 

\item

Let us break $U(N) \to U(N-K) \times U(K)$ by a Higgs vev, keeping
the electric flux in the $U(N-K)$ part. We expect that the low 
energy dynamics of $U(K)$, for a large Higgs vev,  should be
given by the Matrix strings \cite{dvv}, i.e. by an orbifold $Sym(R^8)^K$
perturbed by a dimension 3 operator which depends on the Yang Mills
coupling but not on $N$ or $K$. We check that the low energy interactions
of the closed string modes that couple to NCOS theory have precisely the
correct coupling.

\item

We could consider the gauge theory with $M$ units of electric
flux. This corresponds to having $M$ D-branes in NCOS theory.
We check that the energetics are as expected. 



\end{itemize}
 
After completing these checks of the duality,
we make some remarks on the relation of this NCOS theory 
and the Matrix theory, where the $U(1)$ flux is related to D0 branes. 
In fact we show that NCOS theory in 1+1 dimensions 
is related to DLCQ of IIA string theory in the presence of D0 branes.
We also make some remarks on the high-energy and finite
temperature behaviors of the NCOS theory.

\section{The duality}

In this section we summarize some aspects of the duality 
between $1+1$ dimensional NCOS theory and the $U(N)$ SYM \cite{GMSS}. 

Non commutative open string theory (NCOS) is 
 the theory that results from introducing 
a Moyal phase $\theta = 2 \pi \alpha'_{e}$ in the usual open string
field theory with parameters $\alpha'_e , ~G_0^2$ \cite{SST,GMMS}. 
It is useful to view this theory as arising from a decoupling limit of
usual string theories. These decoupling limits always involve 
$ g_s \to \infty$ so that to properly analyze them we should go to 
S-dual descriptions. {}From this dual description we find that in the 
1+1 dimensional case the decoupling limit reduces to the 
usual decoupling limit for $N$ D-branes in the presence of a weak electric
field \cite{GMSS} which gives a $U(N)$ super Yang Mills theory with
one unit of electric flux. Here   
 $N$ is the number of units of the electric flux
on the D1 brane in the picture where we define the NCOS limit. 
The subscipt on 
$\alpha'_e$ is introduced to distinguish it from the $\alpha'$ of
the theory that has been used to define a decoupling limit:
$\alpha' = \alpha_e' (1-F^2)$, where $F=F_{01}$ is the electric field
on the original D1-brane.

Now let us find the relation between the parameters $\alpha'_e, ~G_0^2$
of the NCOS and the parameters $N, g^2_{YM}$ of the gauge theory.
In order to do this it is convenient to go to a parent string theory
where the $N$ D-branes live and take then the limit $\alpha' \to 0$. 
If we analize this limit in the S-dual theory we find the NCOS limit.
In \cite{GMSS} this limit was discussed. We note here that we
are interested in defining a decoupling limit 
of the theory of a D1 brane in the presence of $N$ units of $U(1)$ flux.

The relation between $\alpha'_e$ and the Yang-Mills coupling
can be found following the steps in the duality to be (note
that our convention for $g_{YM}^2$ differs by a factor of $2\pi$
from the standard one): 
\be 
\label{alphap}  g^2_{YM} = { N^2 \over \alpha'_e} 
\ .\ee
The square of the open string coupling becomes \cite{GKP,GMSS}
\be \label{coupling}
G_0^2 = {1 \over N}
\ .\ee
 
More generally we could consider $M$ coincident D-branes in 
NCOS theory. Now the relation between the electric field on
$M$ D1-branes and the number $N$ of flux unit is modified compared
to the $M=1$ case \cite{CK,IKtalk}.
The relation between parameters is found  to be
\be \label{mult}
g^2_{YM} = { N^2 \over M^2 \alpha'_e}\ , ~~~~~~~~~~~~
G_0^2 = {  M \over N} \ .
\ee
where the dual theory is $U(N)$ SYM with 
$M$ units of flux. It is interesting that now
$G_0^2$ may be an arbitrary fraction. Properties of the theory 
should depend
sensitively on the value of the numerator and denominator: they
should vary discontinuously as a function of $G_0^2$.

In the rest of the paper we check some of the predictions of
this duality.

\section{ Free the U(1) ! }

The $U(1)$ modes of the gauge theory are dual to
the massless modes in NCOS: the scalars in the vector 
representation of $SO(8)$ and their superpartners.
One of the striking predictions of the duality is that the massless
open string modes should decouple. 
This prediction is even more striking when we combine it with the 
statement that all the tree level amplitudes in the NCOS theory
are the same as in the usual open string theory except for some 
Moyal phases. Thus, one might conclude that there must
be a contradiction in the above statements since the tree level 
amplitudes involving massless string states are certainly non-vanishing
in the usual string theory. 
What is important to realize, however, is that the statement
about the Moyal phases is a statement about {\it ordered} 
correlation functions of vertex operators which correspond to
specific Feynman diagrams. The complete amplitudes are given by 
suming over all orderings and thus the phases introduce the
crucial difference. 

Consider the simple case of a four point function. The
relevant disk diagram decomposes into 6 possible orderings. 
There are 2 orderings for 3 of the vertex operators which are
fixed at $0,1,\infty$ and the fourth
can be integrated over $( -\infty ,0)$, $(0, 1)$ or $(1, + \infty)$. 
These three integrals are nonzero individually but their sum
turns out to vanish!
We perform these calculations explicitly 
in Appendix A. 

In the rest of this section we will present a general argument
for why these amplitudes vanish. This 
is due to the precise value of the Moyal phase in NCOS and it
would not be true away from the $F =F_{cr}$ limit. 
In this limit the correlation function of world sheet fields
at the boundary is given by 
\be \label{correl}
\langle X^+(t) X^-(0) \rangle = 4 \alpha'_e ( \log |t|  + i {\pi\over 2}
{\rm sign}(t) )
\ee
where $X^\pm = X^0 \pm  X^1$. 
The crucial observation is that we can think  of the right
hand side of (\ref{correl}), up to an irrelevant constant, 
 as $\log(t)$, where $\log(t)$ is a holomorphic function
of $t$ such that when we go from $t>0$ to $t<0$ it produces the 
phase difference $i \pi$ that we see in (\ref{correl}).
There is of course a question of branch cuts and we have to 
choose them in a definite way. 
Let us suppose that we are integrating a given massless
vertex operator over the boundary of a disk. Since it is 
massless the on-shell condition implies $p_+ p_- =0$. 
 Let us assume that $p_- =0$. Then we choose all branch cuts in 
the log in (\ref{correl}) so that they are outside the disk. 
Then we can deform the contour of integration for $t$ from the boundary
of the disk to the interior and make it disappear, so that the
resulting amplitude is zero. Notice that we did not have to assume
that other vertex operators were also massless: the only assumption
was that at least one  is massless and we integrate that one first. 
Notice that this implies that three point amplitudes are also zero
once we add over all orderings since we can take one of the vertex
operators
and integrate it dividing by the  volume of integration, which is
finite if we integrate only one. Alternatively we can argue that
massless 3-point functions vanish by factorizing the 4-point functions.  

For higher loop diagrams it seems that one 
needs to sum over the insertion of the 
massless state on all boundaries of the diagram in order 
to show that the amplitudes vanish. It would be interesting
to analyze it more precisely.

Just in case the reader fears that all amplitudes would be zero we
compute a four tachyon amplitude in bosonic NCOS theory in Appendix
A, and we show that it is non-vanishing. 
In general, purely massive correlators do not vanish, which agrees with
the expectation that $SU(N)/Z_N$ gauge theory has non-trivial dynamics.
However, as we show in Appendix A, all massive correlators fall off faster
at high energies
than in the commutative open string. The 
$2\rightarrow 2$ scattering cross-section
depends on the energy as $E^{-4}$ away from the poles.
This is reminiscent of how it works in the gauge theory: 
$\sigma \sim g^4_{YM}/E^4 $.

It should be noted that our argument about the decoupling
of massless modes may be extended to 
$p+1$ dimensional NCOS theory with $p>1$. The statement is that
the $ U(1)$ modes with momentum purely in the 1 direction (i.e.
the direction of space-time non-commutativity) decouple
from the theory. In the $p=3$ case, analyzed in \cite{GMMS}, 
the NCOS is dual to a space-space non-commutative Yang-Mills theory
with $B_{23} \not =0$. In general,
the $U(1)$ modes do not decouple in a non-commutative theory,
but if we have a mode that does not depend on $x^2$ and $x^3$,
then it does decouple because the $U(1)$
theory in the remaining directions is commutative. 
Therefore, our argument translates into non-trivial checks
of other dualities in \cite{GMSS} as well.

\section{ Higgs branch and closed strings}

In this section we analyze the second prediction of the duality:
the existence of a Higgs branch with the right energetics. 
It is convenient for this discussion to compactify the Yang-Mills
theory, and also the NCOS theory, on a circle. So we compactify the
direction 1 on a circle of radius $R$. 
The $SU(N)$ with one unit of $Z_N$ flux may be Higgsed
to an $SU(N-K) \times U(1)^K $ with one unit of $Z_{N-K}$ flux in
the first factor. The energy 
cost for this can be calculated from the BPS mass formula for $(p,q)$ 
strings in the gauge theory limit. The relevant formula says that
in a $U(N)$ theory with $M$ units of electric flux we have a ground state
energy of the form 
\be \label{bps}
 E(M,N) = {2 \pi R } {  g^2_{YM} \over 4 \pi  } { M^2 \over N} \ .
\ee
Notice that this gives the energy above the 
 $U(N)$ Yang Mills ground state with no flux. 
The energy of the Higgs branch we considered above, for which $M=1$,
 is 
\be
\label{higgs}
E_K = E(1,N-K)  =  { R g^2_{YM}\over 2 } { 1 \over N-K}
\sim 
 { R g^2_{YM}\over 2 }  \left( { 1 \over N } + 
{ K \over N^2 } + { K^2 \over N^3  } + \cdots 
\right)\ .
\ee
On this Higgs branch we also expect to have a continuum of excitations,
corresponding to configurations where the Higgs field 
has some momentum in the transverse directions. There are also massless
excitations propagating along the $U(1)$s. 
Let us consider first the case of $K=1$: just a single $U(1)$. 
Then we expect to find an energy spectrum for this $U(1)$ which to
leading order in $N$ would be 
\be \label{prediction}
E_1 = { R \over 2 \alpha'_e} + { 1 \over  R } 
(  N_L + N_R  + \alpha'_e  k^2/2 )
\ee
where we have used the relation (\ref{alphap}). 
Notice that the particular dependence of (\ref{prediction}) on $N_{R,L}$
is just stating that we have massless free fields. 
There is also 
a level matching condition that $N_L - N_R$ should be an integer.
 Notice that in 
(\ref{prediction}) we have used only the second term in the expansion
\ref{higgs}. 
The first term represents the mass of the electric
flux line itself:
\be \label{fluxten}
(2\pi R) {N\over 4\pi \alpha'_e} =  (2\pi R) {1\over 4\pi \alpha'_e G_0^2}
\ .
\ee
{}From the dependence of the tension on $G_0^2$ we conclude that the flux 
line is dual to a D-string in the NCOS (this is not surprising
because this is precisely the S-duality relation). 
Note, however, that (\ref{fluxten}) is a factor of two smaller than
the naive expectation for the D1-brane tension in NCOS. 
The naive expectation is what we get by just stating that
the mass of brane is given by the lowest term of the non-commutative
analog of the Born-Infeld action \cite{BI,ACNY}. 
On the other hand, if we just stick to the standard string field 
theory methods and we determine the mass of the brane in the 
the manner described below then we get the right value. 
In order to determine the mass of the brane, imagine that we
do not take the complete decoupling limit but that we 
still keep the original $\alpha'$ finite. Then one way of computing 
the mass of the brane would be to consider the string field $\phi$ 
that parametrizes the position of the brane and to write its kinetic
term as $ S = \int dt C (\dot \phi)^2 $. Then we relate $\phi$ to 
the physical displacement $r$ of the brane by considering 
a configuration with two D-branes and relating the mass of the open
string computed in string field theory, which is linear in $\phi$, 
to the mass we expect if we think we have a string stretching 
between the branes. Finally we write the action as $S = \int dt M 
{\dot r}^2/2 $ and we read off the mass. 
The only difference in this procedure 
between the usual case and the case with the Moyal phases is that
the mass of the stretched string as a function of $r$ is different.
In usual open string theory it is $m = { r \over 2 \pi \alpha'}$
while in this case it is 
$m = { r \over 2 \pi \alpha'_{eff} \sqrt{ 1-F^2} }$.
Computing the mass using this method we get a mass that diverges
since it represents the mass of the $(N,1)$ string in the parent
string theory before taking the decoupling limit. If we subtract
from this mass the mass of the $N$ D1 branes then we indeed
get (\ref{fluxten}). It would be nice to derive the D-brane mass
formula directly in the NCOS theory. 
We believe that the arguments relating the cubic string field theory
action to the Born Infeld  action should fail in the NCOS theory. In fact, 
we saw that the massless states are free, which is incompatible
with a Born-Infeld action.



\subsection{Closed Strings} 

The factor of 2 discussed above is not the only puzzling factor in 
(\ref{higgs}). The tension of the Higgs branch excitation
calculated in (\ref{prediction}) is $1/(4\pi \alpha'_e)$.
Its scaling allows us to identify it with a closed string wound
around the compact circle, except for a factor of 2.
Thus, the duality predicts that,
even though the masses of open strings are given by the usual formula,
$m^2 = N/\alpha_e'$ ($N$ is the integer oscillator level), the 
compactified NCOS theory contains wound closed strings of 
tension $1/(4\pi \alpha'_e)$, instead of the usual closed strings
of tension $1/(2 \pi \alpha'_e)$. In the following we confirm this
prediction. 

The presence of wound closed strings in NCOS is hardly a surprise:
once we compactify NCOS then 
an open string can stretch around the circle and the two ends
can join with a finite probability, since the open string coupling constant
is finite. Therefore,
the NCOS theory can emit wound closed strings into the bulk.  
Moreover the argument leading to (\ref{higgs}) predicts that these strings
can only wind with one orientation: the winding number $K$ cannot be negative. 
We also expect this from the NCOS theory
since the energy necessary to stretch the string in one direction
is finite while it is infinite in the other direction in the 
scaling limit of \cite{SST,GMMS}. 

In order to show directly that these closed strings are produced
we sketch the calculation of the non-planar one loop diagram.
The compactness of the 1 direction introduces the
crucial modification that allows closed string poles, and also
explains why the tension is twice smaller than in the commutative
case (in \cite{GMMS} it was shown 
that there are no closed string poles in the non-compact 
case). 

We consider a one-loop diagram, which is a cylinder. We
introduce some vertex operators on one boundary and some 
on the other boundary so that
we have some net momentum $p$ flowing along the cylinder from one
boundary to the other. As usual, $p^1$ is quantized: $p^1=n/R$. 
We are interested in poles arising from the large $s$ region, where
$s$ is the proper time along the closed string channel. 
These poles must be associated with on-shell closed strings.
In presence of the NS-NS field $B_{01}$,
the on-shell condition for closed strings with $n$ units of momentum
and $m$ units of winding can be derived with standard methods (see,
for example, \cite{Polch}):
\be
-\alpha' (p_0)^2 - 2 p_0 B_{01} mR + {(mR)^2\over \alpha'}
(1- B_{01}^2) + \alpha' (n/R)^2 + 2 (N_L + N_R) + \alpha'_e k^2 =0
\ ,
\ee
where $k$ is the transverse momentum.
These strings do not acquire infinite
energy in the scaling limit $\alpha'= \alpha'_e (1-B_{01}^2) \rightarrow 0$
where $\alpha_e'$ is held fixed. 
Terms quadratic in $p_0$ and $n/R$ vanish, and we obtain a mass formula
of non-relativistic type:
\be \label{onshell}
p_0 = { m R \over 2 \alpha'_e } +    { \alpha'_e 
\over 2 m R} k^2  + {  N_L + N_R \over m R} 
\ee
together with the standard level-matching condition $N_L - N_R= mn$.
This formula is in {\it precise} agreement with the prediction of
the duality (\ref{prediction}). Especially striking is the fact
that that the tension of the wound strings is indeed
$1/(4\pi \alpha'_e)$.
It is crucial that the quadratic dependence on $p_0$ cancels
since it implies that $m>0$ for positive energy states, so that 
strings can wind only in one sense.
Notice also the 
appearance of a continuum associated to the momentum $k$ in the transverse
directions.

\begin{figure}[htb]
\begin{center}
\epsfxsize=1.5in\leavevmode\epsfbox{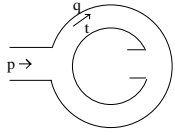}
\end{center}
\caption{
Nonplanar open string diagram. In open string field theory we would 
build it from the cubic vertex and we would consider states carrying 
momentum $q$ and $q+p$ along the loop.
}
\label{nonplanar}
\end{figure}

We can also see the existence of the closed string poles directly in 
the open string interpretation of this diagram, in the 
spirit of the explanation in \cite{GMMS}. 
The idea is to think about the non-planar diagram in open string
field theory where one would have to integrate over the momentum
propagating along the loop (see Figure 1). Then the only difference between
the NCOS diagram and the usual diagram is the presence of the
phase $e^{ 2i p\wedge q}$ where $p\wedge q = \pi \alpha'_e( p_0 q_1 -
p_1 q_0)$, $p$ is the momentum flowing through the closed
string channel and $q$ is the momentum running along the loop.
But now, since momenta are quantized we get that $q_1 = n/R$. 
So we integrate over $q_0$ and sum over $n$. 
The rest of the diagram has the same value as in the usual case. 
For simplicity we set $p_1=0$. 
The propagators can be exponentiated and we have to compute something
of the form
\be
\int dq_0 \sum_n  e^{ i 2 \pi \alpha'_e p_0 n/R } e^{ - 2 \pi \alpha'_e 
(-q_0^2 + (n/R)^2 ) t + \beta p_0 q_0 } \ .
\ee
The integral over $q_0$ is of the usual form and will produce the usual
${\rm exp}( + s {\alpha'_e \over 2}  p_0^2 )$ in the closed string channel 
(after analytic
continuation). The sum over $n$ is the interesting part.
By Poisson resummation we will get
\be
\sum_m e^{ -s { \alpha'_e \over 2} | mR/\alpha'_e - p_0|^2 }
\ .\ee
We now have to multiply this times the result of the integral over $q_0$
times the rest of the string diagram which for small $t$ or large $s$ 
has the standard expansion in terms of closed strings coming from 
modular transforming the small $t$ diagram. 
Putting it all together we have 
\be
\sum_m  \exp\left\{ -s \left({ \alpha'_e \over 2}( |mR/\alpha'_e -p_0|^2 -p_0^2 +
k^2 )  + 2 N \right) \right\} 
\ .\ee
We see that the quadratic part in $p_0$ cancels and then there is 
a pole at 
\be
p_0 = { m R \over 2 \alpha'_e } +   {  \alpha'_e 
\over 2 m R} k^2  + { 2 N \over m R} 
\ee
where $N = N_L = N_R$ which are equal because we assumed $p_1=0$.
In the case of non-vanishing $p_1$ the on-shell condition generalizes to
(\ref{onshell}).

It is interesting to consider the interaction of these closed 
strings. {}From the gauge theory we expect that their low energy 
interactions are described by a orbifold theory 
$Sym(R^8)^K$ deformed by a dimension 3 operator with a coefficient
of the order of  $ {1 \over g_{YM}}$ \cite{dvv}. {}From the string theory 
description we expect the same but with a coefficient 
$G_0^2  \sqrt{\alpha'_e}$. Fortunately, using (\ref{alphap}),
(\ref{coupling}) we find that the two coincide. 

Now let us turn  to the interpretation of the third term in (\ref{higgs}).
The  NCOS action will contain a term of the form
\be
S = { R \over  \alpha'_e G_0^2 }{ 1\over 2}   (2 \pi \alpha'_e)^2 \dot A_1^2 
\ .\ee
For configurations where $A_1 \sim A_1 + 1/R $
depends only on time,
the Moyal phases do not appear. 
As in the usual discussion, if we consider a configuration with 
$K$ units of momentum conjugate to $A_1$, then the energy, 
expanded to quadratic order, will be
\be
E = { R \over  \alpha'_e} { G_0^2  K^2 \over 2}
\ .\ee
We see from (\ref{alphap}), (\ref{coupling}) that we reproduce the 
third term of (\ref{higgs}). 
Since the total winding number for fundamental strings is being kept
fixed we find that this $A_1$ field will have to carry momentum 
precisely equal to the number of strings that have been emitted into the
bulk. 
 
\subsection{Finite Temperature}

Instead of compactifying the $x^1$ direction we could consider
the Euclidean theory and compactify $x^0$ on a circle of radius $R$.
This describes NCOS at temperature $1/(2\pi R)$. Considering closed strings
wound around the Euclidean time direction, we find the dispersion relation
\be
\alpha' (p_1)^2 + 2i p_1 B_{01} mR + {(mR)^2\over \alpha'}
(1- B_{01}^2) + \alpha' (n/R)^2 + 2 (N_L + N_R -1) + \alpha'_e k^2 =0
\ .
\ee
Notice the tachyonic shift now present in the oscillator energies
\cite{AW}.
In the scaling limit we have
\be 
-i p_1 = { m R \over 2 \alpha'_e } +    { \alpha'_e 
\over 2 m R} k^2  + {  N_L + N_R - 1\over m R}\ . 
\ee
As the temperature is increased, a massless closed string state
appears at $T=1/(\pi\sqrt{8\alpha'_e})$. This is precisely the Hagedorn
temperature expected from studying the open string density of states.
It would be interesting to study the connection between the closed 
strings and the NCOS Hagedorn transition in more detail.

\subsection{Coincident D-branes}

Here we consider the case of $M$ coincident 
 D-branes in NCOS theory.
The $U(1)$ part of the massless degrees of freedom is free as before.
On the other hand we have other nearly   massless excitations.
These excitations might not be   truly massless once we include
the effects of the interactions.
This should be dual to $U(N)$ Yang-Mills theory with $M$ units of flux.
In the Yang Mills theory we expect the energy of this
configuration to be proportional to 
\be
{  R g^2_{YM} {M^2 \over 2 N}}
\ .\ee
Using (\ref{mult}) we find that
this agrees with the expected mass of $M$ D-branes in NCOS theory,
up to a factor of two mentioned earlier. 

A simple observation is that the energy cost to separate them in 
NCOS theory is zero to leading order in $ M/N = G_0^2$. 
If $N$ and $M$ are
 coprime we expect that this marginal direction should be
completely lifted. 
In this case we expect from the field theory 
 that the energy cost to separate them
is of order $1/N^3 \sim G_0^2/\alpha'_e$ (up to factors of order $M$).
 This is an effect that 
should arise at one loop. Naively the one loop diagram in NCOS
is the same as in the usual string theory, so this effect would vanish.
What should probably happen is that the mechanism that ensures that 
$G_0^2$ is quantized will in this case produce a $F$ flux on the branes
that would give rise to the energy difference. It would be interesting
to understand the physics quantizing the coupling.

\section{Conclusions}

In this paper we have carried out some consistency checks on the
duality conjectures involving non-commutative open string theory
(NCOS). We have mainly focused on the 1+1 dimensional case which is the
only one where the non-commutativity is consistent with the Lorentz 
invariance (although the parity 
invariance is obviously broken). 
In this case the NCOS is obtained in the limit of critical electric
field along a D-string, and it is believed to be dual to the maximally
supersymmetric 1+1 dimensional $U(N)$ gauge theory with a single flux 
quantum. Thus, 1+1 dimensional NCOS is a new example
of gauge field/string duality \cite{GT,AP}.

In the SYM theory the $SU(N)/Z_N$ dynamics has a mass gap
while the $U(1)$ part decouples at all energy scales.
Perhaps our most striking result is that the dual NCOS respects this
decoupling through the vanishing of all amplitudes involving massless
particles.

We have also considered the NCOS with a compact spatial dimension along
the electric field, and showed that in this case open strings couple to
winding modes of the closed string sector. The mass formula
for such closed strings is in accord with the energetics on the
Higgs branch of the gauge theory: the bulk closed strings are dual to 
individual D-strings which have become liberated
from the $(1,N)$ bound state.

After some number of such closed strings have been liberated, their 
assembly gives rise to a Matrix string theory set-up \cite{dvv}.
Indeed,
 1+1 dimensional maximally supersymmetric SYM
 theory appeared in the Matrix description
of type IIA string theory \cite{dvv}. It was observed there that
one unit of flux in the Yang Mills theory corresponded to 
a D0 brane of type IIA in the infinite momentum frame. 
The large $N$ simplifications that we are seeing 
both in the open string theory
and closed string theory are the simplifications that happen 
for the DLCQ limit of string theory. In particular we are 
analyzing a D0 brane moving at almost the speed of light. 
Indeed after T-dualizing along 1 direction, critical electric field
becomes the velocity equal to the speed of light \cite{CK,SST}.

It is also interesting to notice an analogy between NCOS and the 
SL(2,R) WZW model. In the SL(2,R) WZW model, one has short
strings and long strings \cite{MO}. The short strings are analogous
to the open strings of NCOS and the long strings, which can 
wind only in one direction, are analogous
to the closed strings in NCOS. In both cases the coupling constant
is quantized, at least in the supersymmetric context. 
This is not a coincidence since both theories are describing strings
in critical electric fields or $H$ fields.

It would be  interesting to consider unstable branes in the presence of
electric fields.
There is an important difference between magnetic field and electric
fields on brane-antibrane pair. In the case of magnetic fields
the induced $D(p-2)$ charge in the anti-Dp brane is the opposite to 
that of the $Dp$ brane. So the statement that the tachyon condensation
is the same in the open string theory with Moyal phases as defined
in \cite{Seiberg-Witten} as in the usual string theory is just 
a manifestation of Sen's argument \cite{sen} that brane antibrane 
anihillation is universal.
In the case that we have a constant electric field on the brane
the system has a net fundamental string charge. 
This implies that after tachyon condensation we should be left with
closed strings \cite{sen}. 
It is interesting to notice that we can define the NCOS limit also
for unstable branes like Dp-antiDp branes in string theory,
or D-branes in bosonic string theory. This will
give us an NCOS string theory which has a tachyon. 
Tachyon condensation in NCOS theory is the same as in the usual 
string theory since the tachyon condensate is constant. 
In particular we could consider the case of a 
D1-anti-D1 pair in the NCOS limit. 
This theory is dual to a $U(N)$ theory broken to
$U(N/2)\times U(N/2)$ with opposite $U(1)$ fluxes in each factor.
Then we have a relation between parameters as in 
(\ref{mult}) with $M=2$.
If we naively compute the action at the minimum it seems to be
off by a factor of two from the expected answer. This factor
of two is related to the factor of two present before in the
mass of the D-brane. It would be nice to understand this more 
precisely.

\section*{Acknowledgments}

We are grateful to C. Callan, A. Hashimoto, C. Herzog,
N. Itzhaki, S. Minwalla,
 A. Polyakov, A. Strominger, L. Susskind
and N. Toumbas for useful discussions. 
The research of I.R.K. was supported in part by the NSF grant
PHY-9802484, and the
James S.{} McDonnell Foundation grant No.{} 91-48.  
The research of J.M.\ 
was supported in part by DOE grant DE-FGO2-91ER40654,
 NSF grant PHY-9513835, the Sloan Foundation and the 
 David and Lucile Packard Foundation.

\section{Appendix A}

In this appendix we compute some scattering amplitudes on
D1-branes in the presence of electric fields, following 
\cite{GKP}. We compute them for
a general non-commutativity parameter $\theta$ and then we set
$\theta = 2\pi \alpha_e'$ which corresponds to
the NCOS limit. 

The simplest nontrivial scattering calculation is
the four-point amplitude for massless NS open strings \cite{GKP}:
\begin{eqnarray}
A_{4}(\xi_1,p_1;\xi_2,p_2;\xi_3,p_3;\xi_4,p_4) \sim
 \int \{ d\sigma_1 d\sigma_2 d\sigma_3 d\sigma_4 \} \nonumber \\
\langle \xi_1\inn V_{0}(p_1,\sigma_1)\,
\xi_2\inn V_{0}(p_2,\sigma_2) \, 
\xi_3\inn V_{-1}(p_3,\sigma_3) \,
\xi_4\inn V_{-1}(p_4,\sigma_4) \rangle
\nonumber
\end{eqnarray}
The subindex indicates the picture. 
The vertex operators for scalar particles (the massless
transverse modes of the
string) have the form:
\begin{eqnarray}
V_{-1}^j(p_1,z) = e^{-\phi(z)}\,\psi^j (z)\,e^{ip_1\cdot
X(z)}
\label{vert} \\
V_0^j (p_2,z) =\left(\partial X^j (z)+ip_2\inn \psi(z)\psi^j (z)
\right)\,e^{ip_2\cdot X(z)} \nonumber 
\end{eqnarray}
where $j=2, \ldots, 9$, while the momenta $p^\alpha$ are longitudinal,
$\alpha=0,1$.
$z$ must lie on the real axis. 

The Green function on the boundary:
\be\label{Green}
\langle X^\alpha (\sigma_1) X^\beta (\sigma_2) \rangle =
- 2 \alpha_e' 
\eta^{\alpha\beta} \ln |\sigma_1-\sigma_2|
+ i { \theta \over 2 }
\epsilon^{\alpha\beta} {\rm sign} (\sigma_1-\sigma_2)
\ .\ee
The second term gives rise to phases that depend on the ordering
of the operators.

The $1+1$ dimensional massless kinematics is very restrictive.
In the center of mass frame the momenta are
\bea
p^{\alpha}_1 = \pmatrix { p \cr p\cr}\ , \qquad
p^{\alpha}_2 = \pmatrix { p \cr -p\cr}\ , \qquad
p^{\alpha}_3 = \pmatrix { -p \cr -p\cr}\ , \qquad
p^{\alpha}_4 = \pmatrix { -p \cr p\cr}\ .
\nonumber
\eea
Let us position $V_2$ and $V_4$ at 0 and 1, $V_3$ at $\infty$, and
integrate over the position of $V_1$ from 0 to 1.
This gives the s-channel contribution to the amplitude that was calculated
in \cite{GKP}:
\be
A^s = 
- {G_0^2\over \alpha'_e} \delta^2 (\sum_i p_i)
 \pi s   {\cos ( s { \theta \over 2 \alpha_e'})\over 
\sin(\pi s)} \xi_1\cdot \xi_3 \xi_2 \cdot \xi_4
\ee
with $s$ given by 
(this corrects a numerical factor in (18) of \cite{GKP})
\be
s=-t=
4 p^2 \alpha_e'
\label{kinvar}
\ee


So far we have examined the s-channel amplitude, but it turns out that
the other channels give comparable contributions. They are not
affected by the phases, and their form is
\bea
A^{t,u} \sim s^2 \xi_1\cdot \xi_3 \xi_2 \cdot \xi_4
\Gamma (-u) \left [ {\Gamma(-s)\over \Gamma (1+t)} +
{\Gamma(-t)\over \Gamma (1+s)} \right ]
\eea
In computing this amplitude we imagine that we artificially add some
momentum in the transverse directions as if we were on a higher dimensional
brane, so that $u \not = 0$ and then we take the $u \to 0 $ limit.
The poles in $u$ cancel and we obtain
\bea
A^{t,u} =
{G_0^2\over \alpha'_e} \delta^2 (\sum_i p_i)
 \pi s   {\cos (\pi s)\over 
\sin(\pi s)} \xi_1\cdot \xi_3 \xi_2 \cdot \xi_4
\eea

Adding this to the s-channel, we have the complete 4-point amplitude
\bea \label{main}
A^4 =
{G_0^2\over \alpha'_e} \delta^2 (\sum_i p_i)
 \pi s   {\cos (\pi s) - \cos (s{ \theta \over 2 \alpha_e'})\over 
\sin(\pi s)} \xi_1\cdot \xi_3 \xi_2 \cdot \xi_4
\eea
For $\theta=0$ this has the correct pole structure, with the $t$ and $u$
channels canceling half of the poles in the s-channel amplitude: they
cancel the poles with the residue of the unphysical sign.
Remarkably, for $\theta = 2 \pi \alpha_e' $, 
the amplitude vanishes identically!

We have also calculated some bosonic string 4-point functions
involving the massless scalar particles. The
amplitude for 4 massless scalars turns out to have exactly the same
form as (\ref{main}). For the forward scattering of a scalar off a 
tachyon, we find that the complete amplitude is
\bea
{G_0^2\over \alpha'_e} \delta^2 (\sum_i p_i)
 \pi (s+1)  {\cos [\pi (s+1)] - \cos [(s+1){ \theta \over 2 \alpha_e'}]\over 
\sin [\pi (s+1)]} \xi_1\cdot \xi_2 
\ ,\eea
which again vanishes in the NCOS limit, $\theta = 2 \pi \alpha_e' $.

If all operators are massive, then 
the amplitude no longer vanishes in general.
To check this we have calculated the complete 4-tachyon amplitude on the
bosonic D1-brane:
\bea
A= - {G_0^2\over \alpha'_e} \delta^2 (\sum_i  p_i)
 \pi (s+2)   
{\cos ({\theta \over 2 \alpha_e'} \sqrt{s(s+4)} ) - \cos (\pi s)\over 
\sin(\pi s)}
\eea
Obviously, this no longer vanishes for $ \theta = 2 \pi \alpha_e'$.
 However, in the NCOS limit
the amplitude has much softer $UV$ behavior than the $\theta=0$ amplitude.
Indeed, for large $s$ we may approximate
$$ \cos (\pi \sqrt{s(s+4)} ) \approx \cos [\pi (s+2) - 2\pi/(s+2)]
\ .
$$
Therefore, for large $s$,
$$ A\sim {G_0^2\over \alpha'_e} \delta^2 (\sum_i  p_i)
{\sin [\pi s - \pi/(s+2)]\over \sin [\pi s]}
\ ,
$$
which implies that the residues of the poles at $s=n$ fall off as
$1/(n+2)$ for large $n$. Away from the poles, however, the
amplitude is energy independent at high energies. This should be contrasted
with the commutative open string where this amplitude grows as $E^2$.

This softened UV behavior of NCOS massive amplitudes is general.
If we consider 4 particles of mass $m$, then the amplitude has a factor
$$ \cos (\pi \sqrt{s(s- 4 m^2 \alpha'_e)} ) - \cos (\pi s)
\ ,
$$
which leads to a cancellation for large $s$.

It is interesting that this behavior of 4-point 
amplitudes is similar to the
energy dependence in the perturbative Yang-Mills theory in 1+1 dimensions.
Away from the poles the NCOS amplitude is of order
${G_0^2\over \alpha'_e}$ in the high-energy limit, so that the
$2\rightarrow 2$ scattering cross-section is
\be
\sigma \sim {G_0^4\over (\alpha'_e E^2)^2}
\ .\ee
The perturbative amplitude for 4 W-bosons is energy-independent and of order
$g_{YM}^2$, so that the scattering cross-section is $\sim (g_{YM}/E)^4$.

\end{document}